\documentclass[a4paper,20pt]{article}
    \usepackage[T1]{fontenc}
    \usepackage{graphicx}
    \usepackage{epsfig}
    \usepackage{amsmath}
    \usepackage{amsfonts}
    \renewcommand{\abstract}{}
\begin{document}
\makeatletter
\renewcommand{\@oddhead}{\textit{YSC'14 Proceedings of Contributed Papers} \hfil \textit{Liu Wei, Tang Zheng-Hong, Li Yong-Da}}
\renewcommand{\@evenfoot}{\hfil \thepage \hfil}
\renewcommand{\@oddfoot}{\hfil \thepage \hfil}
\fontsize{11}{11} \selectfont

\title{Evaluation of Optical Magnitude of Deep Space Spacecraft}
\author{\textsl{Liu Wei $^{1,2}$, Tang Zheng-Hong$^{1}$, Li Yong-Da$^{2}$}}
\date{}
\maketitle
\begin{center} {\small $^{1}$Shanghai Astronomical Observatory, Chinese Academy of Sciences \\
$^{2}$Changchun University of Science and Technology, China\\
liuwei@shao.ac.cn}
\end{center}

\begin{abstract}
Optical-electric technology can measure the tangential position
and velocity of spacecraft. To know the feasibility of the use of
optical-electric technology, it is necessary to estimate the
magnitude of spacecraft first. Since the spacecrafts are
non-self-illumination objects, the estimation formulas of the
optical magnitude of spacecraft is constructed according to the
radiation theory and the extra-atmospheric radiant emittance of
the Sun in the visible light wave band. Taking Chang'e-1 as an
example, the magnitude of it in different situations is
calculated.
\end{abstract}

\section*{Introduction}
\indent \indent Because the spacecraft always needs to fly for long
distances, it is very important to monitor it precisely. The
spacecraft has 6 degrees of freedom in the space, three dimension of
position and three dimension of velocity. The mission to measure the
parameters of the orbit of the spacecraft is performed by different
tracing techniques, like radio and optical technologies. There are
three main radio technologies in deep space network, radio
technology which includes radio interferometry, Doppler system with
double frequency, microwave radio radar. The radio interferometry is
to measure the angle and compare the phase of spacecraft. Doppler
system with double frequency is to measure the velocity of
spacecraft. Microwave radio radar is to measure the distance and the
imaging of spacecraft. However optical-electric technology can
measure the tangential position and velocity of spacecraft, which
can provide additional information to trace the spacecraft more
precisely. But the optical-electric technology has some obvious
limitation, it can not work at daytime and under bad weather.

Chang'e-1 is planned to be the first of a series of Chinese
missions to the Moon. The spacecraft will launch in late 2007 on a
CZ-3A booster and orbit the Moon for a year to test the technology
for future missions and to study the lunar environment and surface
regolith. Although Chang'e-1 will be mainly monitored by radio
technology as planed, it is still valuable to analysis the
feasibility to use optical-electric technique as the assistant of
other techniques. To know the feasibility, it is necessary to
evaluate the optical magnitude of the spacecraft in different
situations firstly. This paper will present the initial results of
the estimation of the optical magnitude of Chang'e-1 according to
the radiation theory and the extra-atmospheric radiant emittance
of the Sun in the visible light wave band.

\section*{The ground irradiance of diffuse refection of satellite surface}
\indent \indent Based on the formulas of Plank blackbody
radiation~\cite{LI Shu}, the monochrome radiant emittance of
blackbody radiation is
\begin{equation}\label{eq:eps1}
M(\lambda,\,T_{0})=10^{12}(c_{1}/\lambda^{5})[\,\exp\,(c_{2}/\lambda
T_{0})-1\,]^{-1}
\end{equation}
\begin{figure}
\begin{center}
\includegraphics[height=0.27\textheight,width=0.9\textwidth]{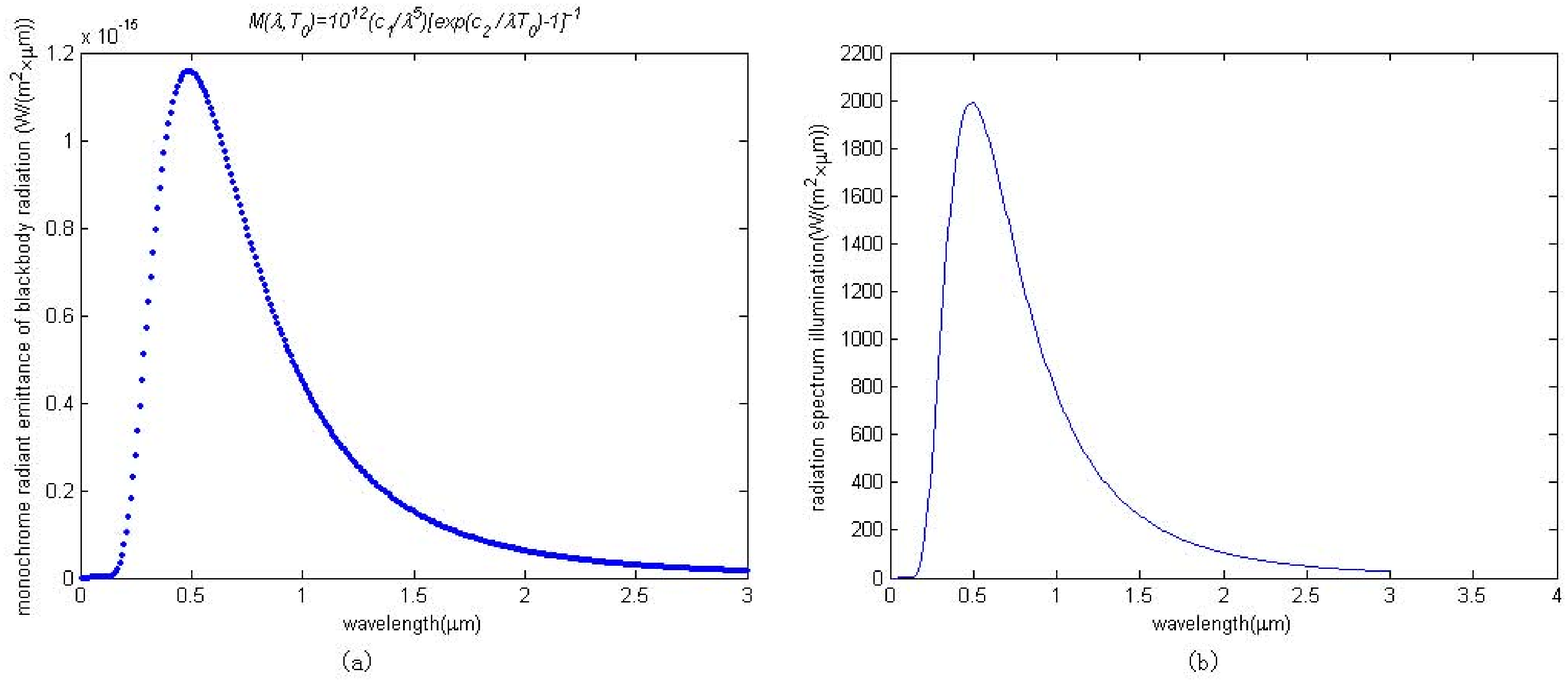}
\caption{The curves change with wavelength}\label{fig:p1_2}
\end{center}
\end{figure}
where $\lambda$ is wavelength (~$\mu$m), $T_{0}$ is blackbody
temperature(K), $c_{1}$=3.742$\times10^{-4}$W$\cdot$~$\mu$m$^{2}$ is
first radiation constant, $c_{2}$=14388$\mu$m$\cdot$ K is second
radiation constant. The unit of the monochrome radiant emittance
$M(\lambda, T_{0}$) is W/(m$^{2}$$\cdot$$\mu$m). Usually solar
radiation can be regarded as blackbody radiation with the
temperature of 5900K, whose monochrome radiant emittance can be
established by (\ref{eq:eps1}).

The solar radiant emittance $M$ and its total radiant flux $\phi$
between $\lambda_{1}$ and $\lambda_{2}$ are
\newcommand{\ud}{\mathrm{d}}
\begin{eqnarray}\label{eq:eps2}
M&=&c_{1}\!\!\int_{\lambda_{1}}^{\lambda_{2}}\!\!\!\lambda^{-5}[\,\exp\,(c_{2}/\lambda
T_{0})-1]^{-1}\ud \lambda\\
\phi&=&4\pi R_{s}^{2}M
\end{eqnarray}
where $T_{0}=5900$K, the solar radius is $R_{s}=6.9599\time10^{8}m$
and the unit of radiant emittance is W/$\mu$ m$^{2}$.

Suppose that the total radiant flux from the Sun is homogeneous in
orientation space, thus the solar luminous intensity from
$\lambda_{1}$ to $\lambda_{2}$ is

\begin{equation}\label{eq:eps3}
I=\phi/4\pi=R_{s}^{2}M
\end{equation}

According to inverse-square law of the distance, the
extra-atmospheric illumination $E_{s}$ in the range of wavelength
between $\lambda_{1}$ and $\lambda_{2}$ is
\begin{equation}\label{eq:eps4}
E_{s}=I/D^{2}=10^{12}R_{s}^{2}c_{1}D^{-2}\int_{\lambda_{1}}^{\lambda_{2}}\lambda^{-5}[\,\exp\,(c_{2}/\lambda
T_{0})-1]^{-1}\ud\lambda
\end{equation}
where $D$ is the actual distance between the Sun and the Earth.
The unit of illumination $E_{s}$ is W/m$^{2}$.
\begin{figure}
\begin{center}
\includegraphics[height=0.25\textheight,width=0.9\textwidth]{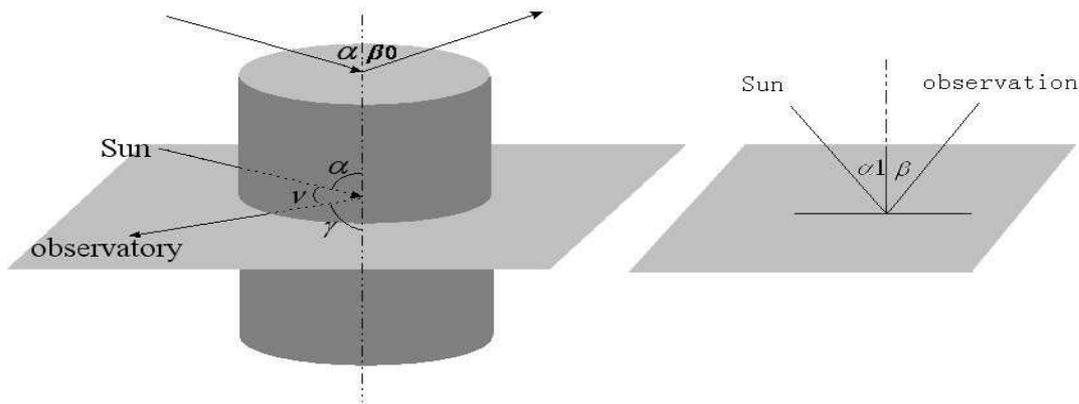}
\caption{The cylinder and plane object}\label{fig:p3}
\end{center}
\end{figure}
The radiation flux of the surface element $\ud$S of the satellite
received from the Sun is
\begin{equation}\label{eq:eps5}
\ud\phi=E_{s}\cos\alpha\ud S
\end{equation}
where $\alpha$ is the angle between the link of the Sun, the
spacecraft and the normal of the surface element $\ud$S of the
spacecraft.

The radiation flux from the surface element $\ud$S of the spacecraft
is
\begin{equation}\label{eq:eps6}
\ud\phi'=\sigma\ud\phi
\end{equation}
where $\sigma$, the diffuse reflectance of the spacecraft, is
independent on the wavelength in the range of visible
light(0.38$\mu$m-0.76$\mu$m). Suppose that $\beta$ is the angle
(rad) between the link of the observer and the spacecraft and the
normal of the surface element $\ud$S, the illumination intensity at
the orientation of which is

\begin{equation}\label{eq:eps7}
\ud I=\cos\beta\ud\phi'/\Omega=\sigma E_{s}\cos\alpha\cos\beta\ud
S/\pi
\end{equation}

So, the ground irradiance of spacecraft diffuse refection from
$\lambda_{1}$ to $\lambda_{2}$ is

\begin{equation}\label{eq:eps8}
E_{m}=\int_{s}L^{-2}\ud I=(\pi L^{2})^{-1}\sigma
E_{s}\int_{s}\cos\alpha\cos\beta\ud S
\end{equation}
where $E_{s}$ is specified by (\ref{eq:eps4}) and L is the distance
between observer and space object and $\tau$, the vertical
atmospheric transmissivity, is independent on the wavelength in the
range of visible light; m, atmospheric mass number, denotes the
ratio of optical thickness between inclined orientation and vertical
orientation.

\begin{equation}\label{eq:eps9}
m=[\cos z+0.150(93.885-z)^{-1.253}]^{-1}
\end{equation}
where z is the zenith angle of spacecraft (rad). When
$z=86^{\circ}$, the error of atmospheric mass number is smaller than
0.1\%.

Most of the spacecrafts are cylindrical, so a cylinder model is
reasonable to be used here. The following formula gives the
irradiance of the cylinder object.

\begin{equation}\label{eq:eps10}
E_{z}=(4\pi
L^{2})^{-1}D_{z}h\sigma\tau^{m}E_{s}\sin\alpha\sin\gamma[\sin\nu+(\pi-\nu)\cos\nu]
\end{equation}
where $D_{z}$ is the diameter of the base surface and $h$ is the
height of cylinder. The position relationship among the angles we
can see from Fig (\ref{fig:p3}). $\alpha$ is the angle of the Sun,
cylinder object and the axial line of cylinder object, $\gamma$ is
the angle between observer, cylinder object and the axial line of
cylinder object, $\nu$ is the angle between the Sun, cylinder object
and observer.

One of the base surface can also be seen.
\begin{equation}\label{eq:eps10}
E_{b}=(4\pi L^{2})^{-1}\sigma\tau^{m}E_{s}\pi
D^{2}\cos\alpha\cos\beta_{0}
\end{equation}
where $\beta_{0}$ is an angle of the base surface towards to the
observation.

We must consider the solar battery planes of the spacecraft, so we
use the plane to stand for the solar battery planes.

\begin{equation}\label{eq:eps11}
E_{p}=(\pi L^{2})^{-1}\sigma\tau^{m}E_{s}S\cos\alpha_{1}\cos\beta
\end{equation}
where $\alpha_{1}$ is incident angle and $\beta$ is reflecting
angle.

Customarily we use equivalent apparent magnitude instead of ground
illumination to figure the luminance of celestial body in
astronomy~\cite{Shen Feng}. The scale is logarithmic, and a
difference of 5 magnitudes means a brightness difference of exactly
100 times. Based on the definition of the equivalent apparent
magnitude, we take the Sun as reference star and its equivalent
apparent magnitude is -26.74. The ground illumination value of the
Sun is $E_{0}$ coming within visible light. The equivalent apparent
magnitude of the spacecraft with the illumination
$E_{m}=E_{z}+E_{b}+2E_{p}$ is written as
\begin {equation}\label{eq:eps11}
m=-26.74-2.5\lg(E_{m}/E_{0}).
\end{equation}

\section*{Simulation}
\begin{figure}
\begin{center}
\includegraphics[height=0.3\textheight,width=0.9\textwidth]{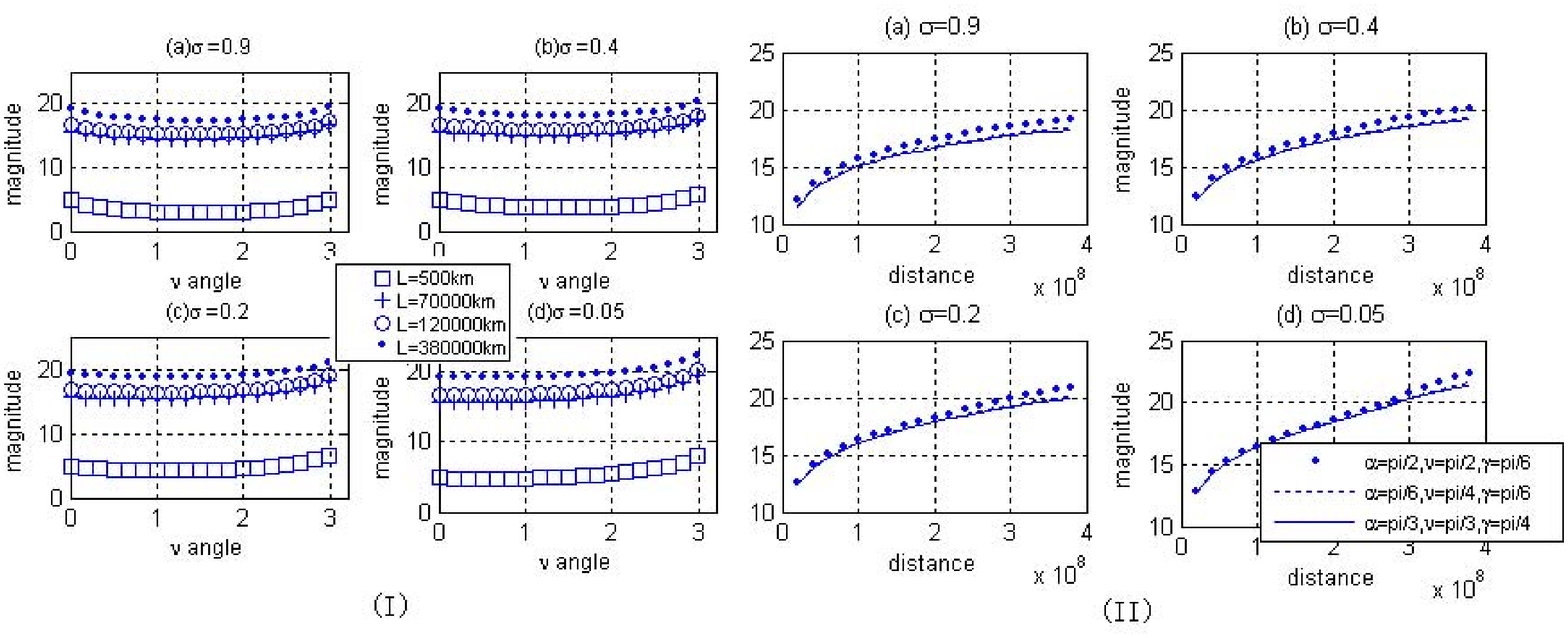}
\caption{The magnitude change curve}\label{fig:The magnitude change
curve}
\end{center}
\end{figure}

\indent \indent We take Chang'e-1 as an example. Its magnitude in
different cases is calculated here. The dimension of Chang'e-1 is
$2000$mm$\times 1720$mm$\times 2200$mm. We consider it as a cylinder
and suppose that its base surface diameter is $D_{z}=2000$mm and its
height is $2200$mm. The average vertical atmospheric transmissivity
is 0.4~\cite{Zhao Yi}. The distance of the spacecraft from lunching
to coming into the orbit of the Moon is from 0 to 380000km. We use
four kinds of color to simulate the surface of the object, whose
surface diffuse reflectance are 0.9, 0.4, 0.2 and 0.05,
respectively. We all use $\pi/4$ as the reflecting angle $\beta$.
The angle relationship of the spacecraft would be told at length in
pictures. We suppose that one of the area of the solar battery
planes is 22.6 $m^{2}$. The diffuse reflectance of them is 0.01. The
angles of $\alpha_{1}$ and $\beta$ are from 0 to $\pi/2$.

In Fig 3(I) y-axis represents magnitude and x-axis represents the
angle of $\nu$ with a variety of $0\rightarrow\pi$. In the four
pictures we employ four different kinds of diffuse reflectance shown
in Fig 3(I). We use some random angles. Dots represents the data of
$\alpha=\pi/4$, $\gamma=\pi/6$. Circles represents the data of
$\alpha=\pi/6$, $\gamma=\pi/3$. Crosses represents the data of
$\alpha=\pi/3$, $\gamma=\pi/3$. Squares represents the data of
$\alpha=\pi/3$, $\gamma=\pi/6$. The four kinds of symbols also
represent different distance. They are 500km, 70000km, 120000km,
380000km, respectively. It is seen that if the distance increases,
the magnitude increases too. But it changes not significantly with
the angle. In Fig 3(II) y-axis is the same as 3(I) and x-axis
represents the distance from the spacecraft to the Earth and varies
from 500km to 380000km. Three symbols stand for different
relationship of the angles. We can see them at length in the
picture. These comparisons suggest that the variation of the
magnitude increases with the distance increasing and decreases with
the diffuse reflectance increasing.
\begin{figure}
\begin{center}
\includegraphics[height=0.4\textheight,width=0.8\textwidth]{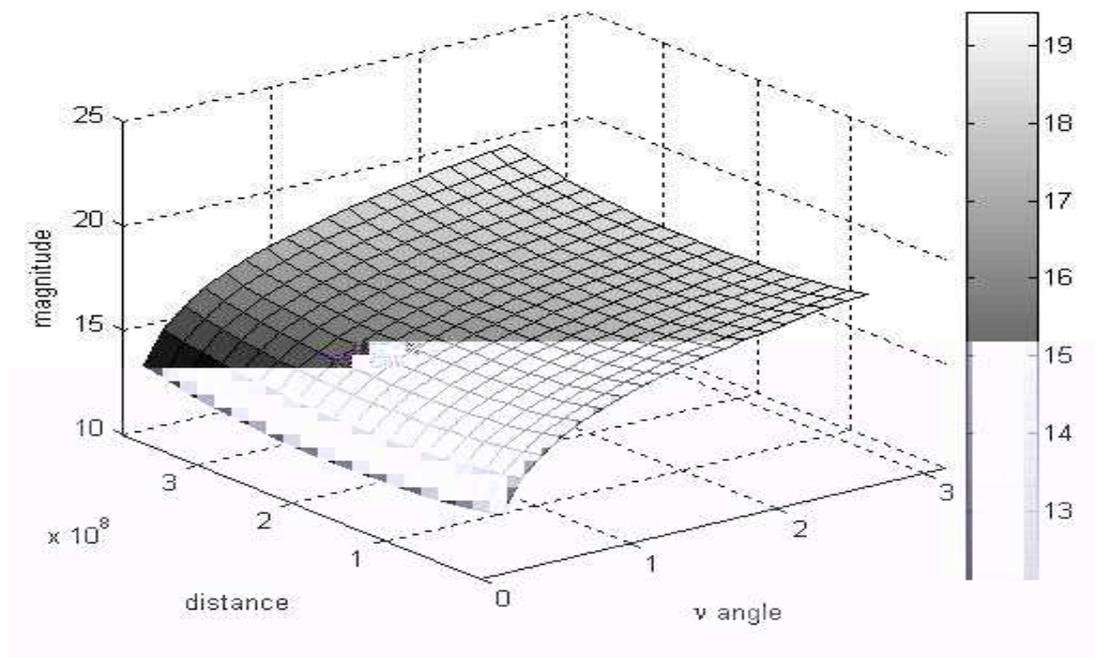}
\caption{The three-dimensional image}\label{fig:3 dimension image}
\end{center}
\end{figure}

We suppose that $\alpha=\pi/3, \gamma=\pi/3, \nu=0\rightarrow\pi$
and $L$ varies from 500km to 380000km in Fig \ref{fig:3 dimension
image}. The three-dimensional image obtained from these data, where
x-axis represent angle of $\nu$, y-axis represents distance and
z-axis represents magnitude. Thus the variation of magnitude of
Chang'e-1 is about 10 magnitude.

\section*{Conclusions}
\indent \indent A model of evaluation the magnitude of cylinder
object is build. From the simulation data, it is found that the
magnitude is dependent on distance, surface diffuse reflectance and
angle. The magnitude varies significantly with distance and surface
diffuse reflectance, but it does not vary with the angle so clearly.
The further the distance is, the larger the magnitude is. It is
different with the relationship between the surface diffuse
reflectance and the magnitude. At last we use the data of Chang'e-1
to calculate, from launching to coming into the orbit of the Moon,
whose variation of magnitude is about 10 mag.

\section*{Acknowledgements}
\indent \indent The author Liu Wei wish to thank her friends Li Yan,
Gao Gang-Jie, Yan Yu-Rong, Tian Yan-Tao, Xiong Wei and Mao Yin-Dun
for their kind help in this work.

\end{document}